\documentclass[conference]{IEEEtran}

\usepackage{cite}
\usepackage{amsmath,amssymb,amsfonts}
\usepackage{amssymb}
\usepackage{algorithmic}
\usepackage{graphicx}
\usepackage{textcomp}
\usepackage{xcolor}
\usepackage{subfigure}
\usepackage{amsthm}
\usepackage{mathrsfs}

\usepackage{pdfpages}
\usepackage{colortbl}
\definecolor{mygray}{gray}{0.85}
\usepackage{array}
\usepackage{caption}
\usepackage{setspace}
\usepackage{booktabs}
\usepackage[linesnumbered,boxed,ruled,commentsnumbered]{algorithm2e}

\usepackage[ruled,linesnumbered]{algorithm2e}

\usepackage{amsmath,bm}
\usepackage{cite}
\usepackage{amsmath,amssymb,amsfonts}
\usepackage{algorithmic}
\usepackage{amsthm}
\usepackage{graphicx}
\usepackage{textcomp}
\usepackage{xcolor}

\makeatletter

\newcommand{\Rmnum}[1]{\expandafter\@slowromancap\romannumeral #1@}
\makeatother

\begin{document}
\begin{spacing}{1.0}
\title{ Reconfigurable Intelligent Computational Surfaces for MEC-Assisted Autonomous Driving Networks}
\vspace{-8mm}



\author{\IEEEauthorblockN{Bo Yang\IEEEauthorrefmark{1}, Xueyao Zhang\IEEEauthorrefmark{1}, Zhiwen Yu\IEEEauthorrefmark{1},  Xuelin Cao\IEEEauthorrefmark{2}, Chongwen Huang\IEEEauthorrefmark{3}, George C. Alexandropoulos\IEEEauthorrefmark{4},  \\ Yan Zhang\IEEEauthorrefmark{5}, M\'erouane Debbah\IEEEauthorrefmark{6}, and Chau Yuen\IEEEauthorrefmark{7}} 
\IEEEauthorblockA{\IEEEauthorrefmark{1}School of Computer Science, Northwestern Polytechnical University, Xi'an, Shaanxi, 710129, China} 
\IEEEauthorblockA{\IEEEauthorrefmark{2}School of Cyber Engineering, Xidian University, Xi'an, Shaanxi, 710071, China}
\IEEEauthorblockA{\IEEEauthorrefmark{3}College of Information Science and Electronic Engineering, Zhejiang University, Hangzhou, China}
\IEEEauthorblockA{\IEEEauthorrefmark{4}Department of Informatics and Telecommunications, National and Kapodistrian University of Athens, Greece} 
\IEEEauthorblockA{\IEEEauthorrefmark{5}Department of Informatics, University of Oslo, 0316 Oslo, Norway}
\IEEEauthorblockA{\IEEEauthorrefmark{6}Center for 6G Technology, the Khalifa University of Science and Technology, Abu Dhabi, UAE} 
\IEEEauthorblockA{\IEEEauthorrefmark{7}School of Electrical and Electronics Engineering, Nanyang Technological University, Singapore}
}

\maketitle

\begin{abstract}
In this paper, we focus on improving autonomous driving safety via task offloading from cellular vehicles (CVs), using vehicle-to-infrastructure (V2I) links, to an multi-access edge computing (MEC) server. Considering that the frequencies used for V2I links can be reused for vehicle-to-vehicle (V2V) communications to improve spectrum utilization, the receiver of each V2I link may suffer from severe interference, causing outages in the task offloading process. To tackle this issue, we propose the deployment of a reconfigurable intelligent computational surface (RICS) to enable, not only V2I reflective links, but also interference cancellation at the V2V links exploiting the computational capability of its metamaterials. We devise a joint optimization formulation for the task offloading ratio between the CVs and the MEC server, the spectrum sharing strategy between V2V and V2I communications, as well as the RICS reflection and refraction matrices, with the objective to maximize a safety-based autonomous driving task. Due to the non-convexity of the problem and the coupling among its free variables, we transform it into a more tractable equivalent form, which is then decomposed into three sub-problems and solved via an alternate approximation method. Our simulation results demonstrate the effectiveness of the proposed RICS optimization in improving the safety in autonomous driving networks.
\end{abstract}


\IEEEpeerreviewmaketitle

\section{Introduction}
\IEEEPARstart{A}{long} with the rapid advancement of wireless technologies, next generation mobile networks will provide low-latency and high-reliability connectivity for intelligent vehicular transportation systems \cite{B5G}. The booming inclusion of artificial intelligence (AI) in industrial automation applications drives the extensive consideration of deep learning techniques for providing intelligence in autonomous driving scenarios by enabling safe navigation. Consider that the on-board sensors may generate huge amounts of multi-modal data, their efficient exploitation for decision making within a limited time becomes a challenge~\cite{NTSB}. In this context, edge intelligence (EI) has become critical, enabling the processing of data uploaded from autonomous vehicles~\cite{YB_wc, EI01}. To further facilitate autonomous driving safety, vehicle-to-infrastructure (V2I) and vehicle-to-vehicle (V2V) wireless links for respectively realizing computation offloading and exchanging critical safety information between vehicles are becoming critical. However, as illustrated in Fig.~\ref{system}(a), V2V receivers `Rx' may suffer from severe co-channel interference caused by neighboring cellular vehicles (CVs), which contributes against driving safety. 

Reconfigurable intelligent surfaces (RISs) play a promising role in improving the quality of wireless links~\cite{RIS00}, thus, they can also assist vehicular communications \cite{RIS01, RIS02}. However, to combat interference at the Rx side of the V2V control link, conventional RISs fall short in providing significant interference suppression needed for autonomous driving~\cite{RIS03}. This happens due to their zero or minimal computing capabilities. To solve this issue, we consider, in this paper, reconfigurable computational intelligent surfaces (RICSs)~\cite{RICS} that are capable of both phase-shifting and amplifying their impinging signals directly in the analog domain, and present a novel framework for their efficient optimization for computation offloading of CVs, while suppressing interference in V2V communication pairs. In contrast to the RIS-aided multi-access edge computing (MEC) approaches in~\cite{RISmec01,RISmec02,yb_twc}, the proposed scheme optimizes an RICS structure to carry out certain computation tasks, rather than solely using the RIS paradigm to improve the offloading link budget via optimized reflective beamforming.  

\theoremstyle{Observation}
\newtheorem{observation}{\textit{Observation}}
\theoremstyle{Lemma} 
\newtheorem{lemma}{{\textit{Lemma}}}
\vspace{-1mm}
\section{RICS-aided Autonomous Vehicular Networks}
\subsection{RICS Modeling}
\begin{figure}[t]
	\centering
	\captionsetup{font={footnotesize }}
	\includegraphics[width=1.0\columnwidth]{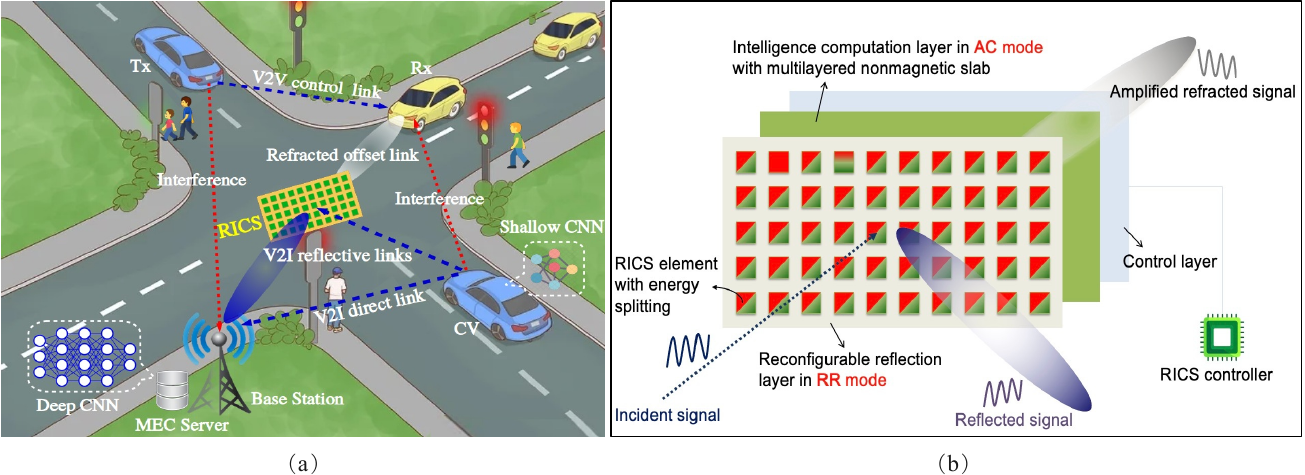}
	\caption{The proposed RICS-aided autonomous driving paradigm is shown in (a), where a V2V Rx suffers severe co-channel interference from neighboring CVs. An optimized RICS can mitigate the interference of the V2V link, while improving the V2I link performance. In (b), the RICS structure is configured as RR+AC mode, being capable to create an ``interference-free zone" via properly configuring the relative permittivity and permeability of the metamaterial included in the intelligence computation layer.}
	\label{system}
\end{figure}

As conceptually sketched in Fig.~\ref{system}(b), RICS is composed of an inner control layer and two functional layers: the \textit{{reconfigurable reflection layer}} and the \textit{{intelligence computation layer}}. In particular, the former layer is configured in a reflection-refraction {(RR) mode} and the latter layer is configured in an {analog computing (AC) mode}~\cite{RICS}. For improving signal coverage, the energy of the signal incident on each element can be split into two parts: a portion of the energy is used for signal reflection, while the remaining of the energy can support signal refraction~\cite{omni,HRIS}. Focusing on this paper's application scenario, i.e., autonomous driving safety, the RICS intelligence computation layer will perform signal amplification on the phase-shifted refracted signals via configuring the tunable metamaterial parameters, targeting interference suppression. 

For the RR mode, let $L$ denote the total number of RICS elements, the energy splitting ratio of the $l$-th element is defined as ${\chi_l} \!\triangleq\! \beta_l^{r}\!:\! \beta_l^{t}, \ \forall l \!\in\! \{1, 2, ..., L\}$, where $\beta_l^{r}$ and $\beta_l^{t}\in [0,1]$ indicate the reflection and refraction amplitude coefficients of each element, respectively, and generally, it holds $\beta_l^{r}+\beta_l^{t}=1$. For the AC mode, inspired by the amplification function achieved in active RISs~\cite{active}, the amplified signal of each $l$-element can be modeled as $y_l={\Psi_l x}+n_s$, where $x$ is the impinging signal, $\Psi_l$ is the amplifying factor of the $l$-element, and $n_s$ indicates static noise. We observe that the signal amplification operation is similar to performing a mathematical operation, with some of them being realized via the metamaterials' metasurface (MS) method in~\cite{metamaterials}.

In this paper, we deploy the MS approach to realize the Fourier transform and its inverse, and the overall system's transfer function. To this end, starting with the original signal $\bm{x}$, the RICS amplifies its amplitude-frequency characteristics by multiplying its Fourier transform with $1/\bm{\Psi}$, where $\bm{\Psi} =\{{{\Psi }_{1}},{{\Psi }_{2}},\ldots,{{\Psi }_{l}}\}$ indicates the vector of signal amplification factors. Then, the inverse Fourier transform is carried out as 
$f(\bm{\Psi}\bm{x})={{\mathcal{F}}^{-1}}\{(\frac{1}{\bm{\Psi}})\mathcal{F}\left[ f(\bm{x}) \right]\}$. In this case, the RICS transfer function is given by $\tilde{\mathop{G}}\,(x)\propto \frac{1}{\bm{\Psi}}$. Considering the width limit of each RICS metamaterial, we need to normalize it to be within the transverse limit of the material, $W$, yielding $G(x)\propto \frac{1}{\bm{\Psi{{x}_{0}}}}$ with ${{x}_{0}}=W/2$. To achieve the desired normalized transfer function proportional to $\frac{1}{\bm{\Psi}}$, the relative permittivity and permeability of the MS method needs to be appropriately configured.

\subsection{System and Channel Modeling}
In the considered RICS-aided autonomous driving scenario of Fig.~\ref{system}, there exist a BS, $M$ CVs that communicate with the BS through V2I links, $N$ V2V communication pairs, and one RICS-like uniform linear array. The CVs can offload their captured images to the MEC server hosted at the BS via V2I links and also share safety information with other CVs via V2V links. The sets of CVs and V2V pairs are denoted by ${\cal M}=\{1, 2, ..., M\}$ and ${\cal N}=\{1, 2, ..., N\}$, respectively. To model the RICS functionality, let $s_l$ denote the signal incident on its $l$-th element. Then, the signals reflected and refracted by this element can be presented as $s_l^{r} = \sqrt{\beta_l^{r}}e^{j \theta_l^r}s_l$ and $s_l^{t} = \Psi_{l}\sqrt{\beta_l^{t}}e^{j \theta_l^t}s_l$ with $\theta_l^r\in(0,2\pi]$ and $\theta_l^t\in(0,2\pi]$ indicating the phase shifts for reflection and refraction, respectively. We also define the reflection and refraction matrices as ${\mathbf{\Phi}}_r={\rm{diag}} \left (s_1^{r}, s_2^{r}, ..., s_L^{r} \right )$ and ${\mathbf{\Phi}}_t={\rm{diag}} \left (s_1^{t}, s_2^{t}, ..., s_L^{t} \right )$,  respectively. To make full use of the available computation resources of the MEC server, current recommendations foresee that V2I communications are assigned the radio spectrum in an time division manner, while V2V communications reuse this spectrum. Let a binary variable $\alpha_{m,n}$ denote the channel sharing between the $m$-th CV and the $n$-th V2V pair. Obviously, CVs will cause interference to V2V pairs if $\alpha_{m,n}=1$, and vice versa.

The wireless channels from each desired CV$_m$ to the RICS, from the RICS to the BS, and from the RICS to the RX of each $n$-th V2V pair, Rx$_n$, are denoted as ${\textbf h}_{m,R} \in \mathbb{C}^{L \times 1}$, ${\textbf h}_{R,B} \in \mathbb{C}^{1 \times L}$, and ${\textbf h}_{R,n} \in \mathbb{C}^{L \times 1}$, respectively. In addition, we model each of the $L$ CV$_m$-RICS-Rx$_n$ links is model as a Rician channel in order to take into account the LoS contribution and the non-LOS (NLoS) multipath components. Therefore, the channel gain is expressed as follows
\begin{equation}\label{h_mr}
{\textbf h}_{m,R}\!=\!   \sqrt{\frac{\kappa_{m,R}}{1\!+\!\kappa_{m,R}}}{\textbf h}_{m,R}^{\rm LoS}+\sqrt{\frac{1}{1\!+\!\kappa_{m,R}}}{\textbf h}_{m,R}^{\rm NLoS}.
\end{equation}
The channel gains of the direct links from each CV$_m$ to BS and from the sender to each Rx$_n$ are denoted by $ h_{m,B}$ and $h_{n}$, respectively. Moreover, the interference channel gain from each CV$_m$ to each Rx$_n$ and from the sender of the $n$-th V2V pair to the BS are given as $h_{m, n}$ and $ h_{n, B}$, respectively. These channels are assumed to be perfectly estimated and quasi-static, hence, remain nearly constant during each transmission time interval. 

The received SINR at the BS from each CV$_m$ and that of each each Rx$_n$ can be obtained as follows
\begin{equation} \label{sinr_m}
\gamma_B^m = \frac{{P_m} \left | h_{m,B} +  {\textbf h}_{R,B} \mathbf{\Phi}_r {\textbf h}_{m,R}  \right |^2 }{\sum_{n=1}^{N}{\alpha_{m,n} P_t} \left | h_{n,B}   \right |^2 + W \xi_0},
\end{equation}
\begin{equation} \label{sinr_n}
\gamma_n= \frac{{P_t} \left |h_{n} \right |^2 }{\sum_{m=1}^{M}{\alpha_{m,n} P_m}  \big|h_{m,n} + {\textbf h}_{R,n}^H \mathbf{\Phi}_t {\textbf h}_{m,R}\big|^2  + W\xi_0},
\end{equation}
where $P_m$ and $P_t$ denote the transmission power of each CV$_m$ and the Tx of each $n$-th V2V pair, respectively. Note that the refracted channel ${\textbf h}_{R,n}^H \mathbf{\Phi}_t {\textbf h}_{m,R}$ is in opposite phase with the direct channel $h_{m,n}$ to cancel the interfering signal at the Rx of the V2V pair. Based on \eqref{sinr_m}, the achievable uplink rate for each CV$_m$ is obtained as $R_B^m=W {\rm log}_2{\left ( 1+ \gamma_B^m \right )}, \ \forall m\in {\cal M}$. The achievable rate for $n$-th V2V pair can be derived accordingly.
\vspace{-4mm}

\subsection{Task and Partial Offloading Modeling}
In this paper, we assume that the CVs capture videos, through their embedded camera sensors, and infer the driving environment in near real-time. We define as an autonomous driving task the task of detecting objects in the environment from the captured images. This task is handled by a pre-trained deep neural network (DNN) deployed at the CVs. For each CV$_m$, each task is characterized by the three-tuple of parameters ${\cal T}_m(s_m, c_m, \sigma_m)$, where $s_m$ [bits], $c_m$ [cycles], and $\sigma_m$ [secs] denote the size image of data to be processed, the total number of CPU cycles required, and the maximum tolerable delay. Without loss of generality, we assume that the DNN models deployed at the CV and BS are have different computational capabilities. For each CV$_m$'s data processing with a given quality $Q$, we assume the inference accuracy be smaller than that achieved by the BS, i.e., $A_m(Q) = \lambda A_B(Q)$ with $0 \leq A_m(Q),A_B(Q)$ and $\lambda \leq 1$.
 
Without loss of generality, a partial offloading model is considered. Specifically, for each CV$_m$, the offloading ratio $\rho_m$ is defined as the portion of the tasks offloaded to the BS. Accordingly, $1-\rho_m$ indicates the ratio of data to be processed locally at CV$_m$. Hence, each task ${\cal J}_m$ of CV$_m$ can be divided into two parts: $\rho_m s_m$ (bits) portion offloaded to the BS and $(1-\rho_m)s_m$ (bits) processed locally. Specifically, the local computation delay of each CV$_m$ is calculated as $\tau^m_{l} = \left ( 1-\rho_m \right ) {c_m}/{f_m}$ with $f_m$ denoting CV$_m$'s computing resources. The total delay introduced by computation offloading is given by $ \tau^m_{o} = \rho_m \left (  \frac{s_m}{R_B^m}+\frac{c_m}{F} \right )$, where $F$ represents CPU resources at the MEC server at the BS. Therefore, the total delay for task ${\cal J}_m$ at CV$_m$ via the considered partial offloading scheme is $\tau_m={\rm max} \{\tau^m_{l},  \tau^m_{o}  \}$, hence, the average inference accuracy can be obtained as  
\begin{equation}\label{Accuracy}
\tilde{A}_m(Q)= \left ( 1- \rho_m \right )A_m(Q) + \rho_m A_B(Q).
\end{equation}

\section{Problem Formulation and Analysis}\vspace{-2mm}
\subsection{Problem Formulation}\vspace{-2mm}
According to $A_m(Q) \!=\! \lambda A_B(Q)$ and~\eqref{Accuracy}, we define the driving safety coefficient as
\begin{equation}\label{saftey1}
{S}_m \overset{\rm def}{\!=\!} \frac{A_B(Q) \left( \lambda \!+\!  \rho_m \left ( 1\!-\! \lambda  \right ) \right )}{{\rm max} \{\tau^m_{l},  \tau^m_{o}  \}}. 
\end{equation}
We next focus on designing the spectrum sharing strategy, the RICS free parameters, and the task offloading ratios to maximize the driving safety coefficients at all CVs under outage probability constraints for the V2V communication pairs. In mathematical terms, we formulate the following optimization problem:
\begin{subequations} \label{binary_original_problem}
\begin{align}
& \;\;\;\; \mathbb{P}: \ \underset{\bm{\alpha}, \mathbf{\Phi}_x, \bm{\rho}}{\rm max}\;\;  \sum\limits_{m\in \mathcal{M}}S_m  \notag \\
& \;\;\;\;\;\;{{s}}{{.t}}{\rm{.}}\;\;\; \; {\rm Pr}\{\gamma_n \leq \gamma_{th}\} \leq P_{\rm out}, \\
& \;\;\;\;\;\;\;\;\;\;\;\;\;\; \;  \alpha_{m,n} \in \{0,1\}, \ \forall m \in {\cal M}, \forall n \in {\cal N},  \\
& \;\;\;\;\;\;\;\;\;\;\;\;\;\;  \sum\limits_{n\in \mathcal{N}} \alpha_{m,n} \le 1, \ \forall m \in {\cal M}, \forall n \in {\cal N},  \\
& \;\;\;\;\;\;\;\;\;\;\;\;\;\; \;  \theta_{l}^r, \theta_{l}^t \in [0, 2\pi),  \ 1 \le l \le L,  \\
& \;\;\;\;\;\;\;\;\;\;\;\;\;\; \; 
\beta_l^{r}+\beta_l^{t}=1, \ 1 \le l \le L, \\
& \;\;\;\;\;\;\;\;\;\;\;\;\;\; \; \rho_m \in [0,1], \ \forall m \in {\cal M},
 \end{align}
\end{subequations}
where $\bm \alpha = \{\alpha_{m,n}, \forall m, n\}$ and  $\bm \rho=\{\rho_1, \rho_2, ..., \rho_m\}$. This problem is a mixed integer non-convex optimization problem (MINLP), which is in general difficult to solve.
\vspace{-2mm}
\subsection{Outage Probability Analysis}
The outage probability constraint in (\ref{binary_original_problem}a) can be rewritten as 
${\rm Pr}\{\gamma_n \leq \gamma_{th}\} = \mathbb{E} \left [u \left(\gamma_{th}- \gamma_n\right)  \right ]$, where $\mathbb{E}[x]$ denotes $x$'s expected value and $u(x)$ is the step function. Although there exist many smooth approximations for the step function, we consider the smooth approximation $\hat{u}_\omega(x)=\frac{1}{1+e^{-\omega x}}$ including with a smooth parameter $\omega$ for controlling the approximation error. By substituting with $\hat{u}_\omega(x)$, we can obtain an approximation of the constraint (\ref{binary_original_problem}a) as follows:
\begin{equation} \label{proof01}
\mathbb{E} \left [\hat{u}_\omega \left(\gamma_{th}- \gamma_n\right)  \right ] \leq P_{\rm out}.
\end{equation} 
 
Due to the concavity of the step function and using Jensens inequality, (\ref{proof01})'s left hand side can be
upper bounded as:
\begin{equation} \label{proof03}
 \begin{aligned}
\mathbb{E} \left [\hat{u}_\omega \left(\gamma_{th}- \gamma_n \right)  \right ] &\leq  
\hat{u}_\omega \left( \gamma_{th}- \mathbb{E} \left[ \gamma_n \right ] \right).
\end{aligned}
\end{equation}
As for $\mathbb{E} \left[ \gamma_n \right ]$, by substituting (\ref{sinr_n}) into (\ref{proof03}), yields:
\begin{equation} \label{proof04}
 \begin{aligned}
\mathbb{E} \left[ \gamma_n \right ] \!&=\! \frac{{P_t} \mathbb{E} \left[ \left |h_{n} \right |^2 \right ] }{\sum_{m=1}^{M}{\alpha_{m,n} P_m} \mathbb{E} \left[\big|h_{m,n} \!+\! {\textbf h}_{R,n}^H \mathbf{\Phi}_t {\textbf h}_{m,R}\big|^2\right ] \!+\! W\xi_0}.
\end{aligned}
\end{equation}

\begin{figure*}[!h]
	\begin{equation}
		\begin{aligned}\label{proof05}
		 &\mathbb{E} \left[\big|h_{m,n} \!+\! {\textbf h}_{R,n}^H \mathbf{\Phi}_t {\textbf h}_{m,R}\big|^2\right ]
		= 1\!+\! \frac{1}{\left(1\!+\!\kappa_{R,n}\right) \left(1\!+\!\kappa_{m,R} \right)} \left(\left|\sqrt{\kappa_{R,n} \kappa_{m,R}} \cdot {\textbf h}_{R,n}^{\rm LoS} \mathbf{\Phi}_t {\textbf h}_{m,R}^{\rm LoS}\right|^2 +N \kappa_{R,n} +N \kappa_{m,R} +N\right).
		\end{aligned}
	\end{equation}
	\rule{\textwidth}{1pt}
\end{figure*}

We now focus our attention on the computation of $\mathbb{E} \left[\big|h_{m,n} \!+\! {\textbf h}_{R,n}^H \mathbf{\Phi}_t {\textbf h}_{m,R}\big|^2\right ]$ in (\ref{proof04}). By assuming that $h_{m,n}$, ${\textbf h}_{R,n}$, and ${\textbf h}_{m,R}$ are independent of each other, (\ref{proof05}) can be derived using the following notations: 
\begin{subequations}
\begin{align}
& {\textbf H}_1 \triangleq \sqrt{\kappa_{R,n} \kappa_{m,R}} {\textbf h}_{R,n}^{\rm LoS} \mathbf{\Phi}_t {\textbf h}_{m,R}^{\rm LoS}, \\
& {\textbf H}_2 \triangleq \sqrt{\kappa_{R,n}} {\textbf h}_{R,n}^{\rm LoS} \mathbf{\Phi}_t {\textbf h}_{m,R}^{\rm NLoS}, \\
& {\textbf H}_3 \triangleq \sqrt{\kappa_{m,R}} {\textbf h}_{R,n}^{\rm NLoS} \mathbf{\Phi}_t {\textbf h}_{m,R}^{\rm LoS}, \\
& {\textbf H}_4 \triangleq  {\textbf h}_{R,n}^{\rm NLoS} \mathbf{\Phi}_t {\textbf h}_{m,R}^{\rm NLoS}.
\end{align}
\end{subequations}

As for the direct V2V link with $h_{m,n}\sim {\cal C N}(0,1) $ and the desired cascaded channels, it holds that $\mathbb{E} \left[\left|h_{m,n}\right|^2\right ] \!=\! 1$, $\mathbb{E}\left[\left|{\textbf H}_1\right|^2 \right ]\!=\!\left|{\textbf H}_1\right|^2$, $\mathbb{E}\left[\left|{\textbf H}_2\right|^2 \right ]\!=\!N \kappa_{R,n}$, $\mathbb{E}\left[\left|{\textbf H}_3\right|^2 \right ]\!=\!N  \kappa_{m,R}$, and $\mathbb{E}\left[\left|{\textbf H}_4\right|^2 \right ]\!=\!N$. By substituting (\ref{proof05}) into (\ref{proof03}), we have 
\begin{equation} \label{proof08}
 \begin{aligned}
\mathbb{E} \left [\hat{u}_\omega \left(\gamma_{th}- \gamma_n \right)  \right ] &\approx \hat{u}_\omega \left( \gamma_{th}- \mathbb{E} \left[ \gamma_n \right ] \right) \\
&=\hat{u}_\omega \left( \gamma_{th}-\tilde{\gamma}_{n}(\bm{\alpha}, \mathbf{\Phi}_x) \right),
\end{aligned}
\end{equation}
where $\tilde{\gamma}_n (\bm{\alpha}, \mathbf{\Phi}_x) = \mathbb{E} \left [ \gamma_n \right ]$ with $\mathbb{E} \left[ \left |h_{n} \right |^2 \right ]=1$.

By using (\ref{proof01}) and (\ref{proof08}), the constraint (\ref{binary_original_problem}a) can be rewritten as $
\hat{u}_\omega \left( \gamma_{th}-\tilde{\gamma}_{n}(\bm{\alpha}, \mathbf{\Phi}_x) \right) \leq P_{\rm out}$. 
By using $\hat{u}_\omega(x)$'s form and deploying an inequality transformation, we have 
\begin{equation} \label{proof02}
\tilde{\gamma}_n (\bm{\alpha}, \mathbf{\Phi}_x) \geq \gamma_{th}+\frac{1}{\omega} {\rm ln} \left ( \frac{1}{P_{\rm out}}-1 \right )\triangleq \tilde{\gamma}_c.
\end{equation}

Next, to make this problem more tractable, 
we first relax the binary variables in (\ref{binary_original_problem}b) into continuous variables, yielding the following problem:
\begin{subequations} \label{equivalent_problem}
\begin{align}
& \;\;\;\; \tilde{\mathbb{P}}: \ \underset{\bm{\alpha}, \mathbf{\Phi}_x, \bm{\rho}}{\rm max}\;\;  \sum\limits_{m\in \mathcal{M}} S_m  \notag \\
& \;\;\;\;\;\;{{s}}{{.t}}{\rm{.}}\;\;\; \; 0 \leq \alpha_{m,n} \leq 1, \ \forall m \in {\cal M}, \forall n \in {\cal N},  \\
& \;\;\;\;\;\;\;\;\;\;\;\;\;\;  (\ref{binary_original_problem}c) - (\ref{binary_original_problem}f), (\ref{proof02}). \notag
\end{align}
\end{subequations}

Using linear relaxation, $\tilde{\mathbb{P}}$'s objective usually provides an upper bound for $\mathbb{P}$'s objective. Note that, in $\tilde{\mathbb{P}}$, there are three optimization variables: the spectrum sharing strategy $\bm{\alpha}$ for V2V links, the RICS reflection matrix $\mathbf{\Phi}_x$, and the offloading strategy $\bm{\rho}$. However, problem $\tilde{\mathbb{P}}$ is a non-convex optimization problem. Due to the coupling of these variables, it is, in general, difficult and almost prohibitive to find the globally optimal solution. Motivated by this, we next develop efficient algorithms to find a high-quality suboptimal solution.

\vspace{-1mm}
\section{Proposed RISC-Aided Inference Solution}
After the analysis in the previous section, the original problem can be rewritten as the equivalent problem $\tilde{\mathbb{P}}$ in (\ref{equivalent_problem}). Due to the non-convexity of the optimization problem as well as the coupling between the three variables, it is difficult to obtain the global optimal solution directly. Therefore, we consider splitting the problem $\tilde{\mathbb{P}}$ into three subproblems and using the alternating optimization method.
\vspace{-2mm}
\subsection{Optimization of the Task Offloading Ratio}
We observe that the original problem is a fractional programming (FP) problem. Suppose the channel-sharing policy $\bm{\alpha}$ and the reflection and refraction coefficient matrix ${\mathbf{\Phi}}_x$ are given, due to the non-negativity of the molecules in (\ref{saftey1}), we introduce the quadratic transform algorithm~\cite{FP_Quad} to convert the original optimization to a linear form. 

Here, we introduce a coefficient $\bm{\mu}$ and then rewrite the programming problem as
\begin{subequations} \label{Quad}
\begin{align}
& \text{        }\underset{\bm{\rho} ,\bm{\mu} }{\mathop{\max }}\,\sum\limits_{m\in \mathcal{M}}{\left(2\mu \sqrt{(1-\lambda ){{A}_{B}}(Q){\bm{\rho }}+\lambda {{A}_{B}}(Q)}-{{\mu }^{2}}{{\tau }}(\bm{\rho })\right)} \nonumber \\ 
&{{s}}{{.t.}}\;\;\;\;\;\;\;\;(\ref{binary_original_problem}f).\nonumber
\end{align}
\end{subequations}

Let $\bm{\mu}^* = \arg\max f(\bm{\rho}, \bm{\mu})$, then we have $f(\bm{\rho},\bm{\mu}^*) = \bm{S}$. Since the function $f(\bm{\rho}, \bm{\mu})$ is convex for the fixed $\bm{\rho}$ and $\bm{\mu}$, we perform the convex optimization of $\bm{\mu}$ through the quadratic transformation algorithm, leading to iterative updates of $f(\bm{\rho}, \bm{\mu})$ and ultimately achieving convergence for the global optimal solution of the FP problem. 




\subsection{Optimization of the Spectrum Sharing Strategy}
When  ${\mathbf{\Phi}}_x$ and $\bm{\rho}$ are fixed, we perform alternating optimization to obtain the optimal spectrum sharing strategy $\bm{\alpha}^*$. The original problem (\ref{binary_original_problem}) can be reformulated as
\begin{subequations} \label{reformulate}
\begin{align} 
& \mathbb{P}:\;\;\;\;\;\;\underset{\bm{\alpha}}{\mathop{\max }}\,\text{  }\sum\limits_{m\in \mathcal{M}}{{{S}_{m}}} \nonumber \\ 
&{{s}}{{.t}}{\rm{.}}
 \text{       }\;\;\;\;\;{{\tilde{\mathop{\gamma }}\,}_{n}}(\bm{\alpha} )\ge {\tilde{\gamma }_{c}}, \\
&\;\;\;\;\;\;\;\;\;\;\;\; (\ref{equivalent_problem}a), (\ref{binary_original_problem}c). \nonumber
\end{align}
\end{subequations}

By observing (\ref{saftey1}), we find that the numerator is fixed once $\bm{\rho}$ is given. For the denominator part, the calculation of the local delay $\tau^m_{l}$ does not involve $\bm{\alpha}$ while only the ${R_{B}^{m}}$ is a function of $\bm{\alpha}$. Therefore, according to the harmonic average, the problem (\ref{reformulate}) can be rewritten into $\underset{\bm{\alpha}}{\mathop{\min }} \sum\nolimits_{i=1}^{m}{{{\tau }_{i}}}$, which, however, is still non-convex due to the coupling between the variables involved. To deal with this, we use the log-sum-exponential function to get an approximation of the max function, i.e., $\max \{{{x}_{1}},{{x}_{2}},...,{{x}_{n}}\}\le \log (\sum\limits_{i\in {N}}{{{e}^{{{x}_{i}}}}})$.
So the problem (\ref{reformulate}) can be further rewritten as the form of log-sum-exponential, i.e.,
\begin{equation} \label{tau}
\underset{\bm{\alpha}}{\text{min }}\sum\limits_{m\in \mathcal{M}}{\tau _{m}^{ub}}=\underset{\bm{\alpha}}{\text{min }} \sum\limits_{m\in \mathcal{M}}{\log \left( {{e}^{\tau _{l}^{m}}}+{{e}^{\tau _{o}^{m}}} \right)}.
\end{equation}

\begin{figure*}[t]
\begin{equation} \label{SCA_Rm}
\begin{split}
{\mathop{\tilde{R}_{B}^{m}}}^{(k)}\,={{p}_{m}}({\bm{\alpha}}^{(k)})-\left({{q}_{m}}({\bm{\alpha}^{(k)}})+\sum\limits_{n\in \mathcal{N}}{{{\left. \frac{\partial {{q}_{m}}(\bm{\alpha})}{\partial {{\alpha}_{m}}} \right|}_{\bm{\alpha}={{\alpha}^{(k)}}}}}\left({{{\tilde{\alpha}}\,}_{m}}-{\tilde{\alpha}}\,_{_{m}}^{(k)}\right)\right).
\end{split}
\end{equation}
\rule{\textwidth}{1pt}
\end{figure*}
\vspace{-2mm}
At this point, we only need to ensure that $\tau^m_{o}$ is convex. Considering that the inverse of the concave function is convex, so we need to figure out the concave property of $R_{B}^{m}$. Specifically, we first expand $R_{B}^{m}$ by using the difference of concave functions (DC), i.e., 
\begin{equation}
\begin{split}
R_{B}^{m}(\bm{\alpha})=\underbrace{W{{\log }_{2}}\left({{\Xi }_{1}}{{\alpha}_{m,n}}+\sum\limits_{n\in \mathcal{N}}{{{\Xi }_{2}}{{\alpha}_{m,n}}}+W{{\xi }_{0}}\right)}_{p_{m}(\alpha )}
-\\\underbrace{W{{\log }_{2}}\left(\sum\limits_{n\in \mathcal{N}}{{{\Xi }_{2}}{{\alpha}_{m,n}}}+W{{\xi }_{0}}\right)}_{q_{m}(\alpha )}, \notag
\end{split}
\end{equation}
where ${{\Xi }_{1}}={{P}_m}|{h_{m,B}}+{{\textbf h}_{R,B}}{\bm{\Phi }_{r}}{{\textbf h}_{m,R}}{{|}^{2}}$, and ${{\Xi }_{2}}={{P}_t}|{h_{n,B}}{{|}^{2}}$.

Obviously, the former part is concave while the latter is convex. Here we consider using the successive convex approximation (SCA) technique at each iteration to approximate the original non-convex problem at a given local point. In the $k$-th iteration, we define ${\bm{\alpha}^{(k)}}\!=\!\{{\alpha}_{m,n}^{(k)},\forall m\}$ as the expansion of the specific point. Since the first half is concave, the DC variant of the problem is still not directly solvable, we approximate it as a convex function with first-order Taylor expansion at a given point. Therefore, an upper bound on the objective function is achieved to obtain the optimal solution. We denote the first-order Taylor expansion of ${{q}_{m}}(\bm{\alpha})$ at the given point ${\tilde{\bm{\alpha}}}$ in the $k$-th iteration as $q_{_{m}}^{(k)}(\bm{\alpha})$,
by substituting into the original equation of ${R}_{B}^{m}$, we can obtain (\ref{SCA_Rm}). In this case, the problem (\ref{tau}) can be further rewritten as 
\begin{align} \label{alpha}
 &\underset{\bm{\alpha}}{\mathop{\min }}\;\;\;\;\;\sum\limits_{m\in \mathcal{M}}{\tau _{m}^{ub}}  \\ 
 &{{s}}{{.t}}{\rm{.}}
 \;\;\;\;\;\;\;(\ref{equivalent_problem}a),  (\ref{reformulate}a), (\ref{binary_original_problem}c). \nonumber
\end{align}

At this point, the objective function and the constraint are convex, so we can solve them by using the standard convex optimization solver such as CVX.
\vspace{-2mm}

\subsection{Reflection and Refraction Matrices Optimization}
After given $\bm{\rho}$ and $\bm{\alpha}$, the original problem is replanned. Currently, our goal is to minimize the sum of the $\bm{\tau}$ and thus our goal is to find an optimal value of ${\mathbf{\Phi}}_x$ that maximizes the sum of $R_{B}^{m}$.
Since both the objective function and the constraint (\ref{reformulate}c) have the variables ${\mathbf{\Phi}}_x$, they can be regarded as a quadratically constrained quadratic programs (QCQP), which can be solved with SDR algorithm.

Considering the optimization variables only appear in the SINR, $\gamma_{B}^{m}$, and the constraints (\ref{binary_original_problem}d)-(\ref{binary_original_problem}e). Since the numerator of the safety coefficient ${S}_{m}$ is also fixed due to the fixation of $\bm{\rho}$, so we rewrite the problem as
\begin{align} \label{phi_reformulate}
& \underset{{{\bm{\Phi} }_{t}},{{\bm{\Phi} }_{r}}}{\mathop{\max }}\,\text{    }\sum\limits_{m\in \mathcal{M}}{{{\log }_{2}}\left(1+\frac{{{P}_{m}}|h_{m,B} \!+\! {\textbf h}_{R,B} \mathbf{\Phi}_t {\textbf h}_{m,R}\big|^2}{\sum\nolimits_{n=1}^{N}{{{\alpha }_{m,n}}{{P}_{t}}|{h_{n,B}}{{|}^{2}}\!+\!W{{\xi }_{0}}}}\right)}  \\
&\;{{s}}{{.t}}{\rm{.}}
\;\;\;\;\;\;\;\;\;\;\;\;\;\;\;\;\;\;\;\;\;\;\;\;\;(\ref{binary_original_problem}d)-(\ref{binary_original_problem}e), (\ref{reformulate}c). \nonumber
\end{align}

To better handle the terms in the SINR expression, $|{h_{m,B}}+{\textbf{h}_{R,B}}{\bm{\Phi}_{r}}{\textbf{h}_{m,R}}{{|}^{2}}$, we introduce the vector ${{\bm{\Theta}}_{x}}\in {{\mathbb{C}}^{L\times 1}}$ as the main diagonal element of $\bm{\Phi }_{x}$, i.e.,  ${{\bm{\Theta}}_{x}}={{[s_{1}^{x},...,s_{L}^{x}]}^{T}}, \forall x\in \{t,r\}$. Then, the auxiliary variables ${{\textbf{h}}_{B}}\in {{\mathbb{C}}^{L\times 1}}$ are introduced and thus we have ${{\textbf{h}}_{B}}={\textbf{h}_{R,B}}\circ {\textbf{h}_{m,R}}$.

Next, we introduce the auxiliary variables $\bm{R}$ and $\bm{v}$,
\begin{equation*}
{\bm{R}_{B}}={{P}_{m}}\left[ 
\begin{aligned}
  & {\textbf{h}_{B}}\textbf{h}_{B}^{H}\;\;\;\;\;{\textbf{h}_{\text{B}}}h_{m,B}^{H} \\ 
 & {\textbf{h}_{B}^{H}}{h_{m,B}}\text{     }h_{m,B}^{H}{h_{m,B}} \\ 
\end{aligned} \right],{{\overset{\_}{\mathop{\bm{v}}}\,}_{x}}=\left[ 
\begin{aligned}
  & {\bm{\Theta}_{x}} \\ 
 & 1 \\ 
\end{aligned} \right],\forall x\in \{t,r\}.
\end{equation*}

Thus, we have $
{{P}_{m}}|{h_{m,B}}+\bm{\Theta} _{x}^{H}{\textbf{h}_{B}}{{|}^{2}}=\overline{\bm{v}}_{x}^{H}{\bm{R}_{B}}{\overline{\bm{v}}_{x}}.$

Due to ${{{{\overline{\bm{v}}_{x}^{H}}}}}{\bm{R}_{B}}{{\overline{\bm{v}}_{x}}}=\text{Tr}({\bm{R}_{B}}{{\overline{\bm{v}}_{x}}}{{{{\overline{\bm{v}}_{x}^{H}}}}})$, we define ${\bm{V}_{x}}={{\overline{\bm{v}}_{x}}}{{{{\overline{\bm{v}}_{x}^{H}}}}}$, which satisfies ${\bm{V}_{x}}\ge 0$ and rank$({\bm{V}_{x}})=1$, then the original problem can be expressed as
\begin{subequations} \label{problem36}
\begin{align}
& \underset{{\bm{V}_{r}},{\bm{V}_{t}}}{\mathop{\max }}\,\text{    }\sum\limits_{m\in \mathcal{M}}{{{\log }_{2}}\left(1+\frac{Tr\left({\bm{V}_{x}}{\bm{R}_{B}}\right)}{\sum\nolimits_{n=1}^{N}{{{\alpha }_{m,n}}{{P}_{t}}|{{h}_{n,B}}{{|}^{2}}+W{{\xi }_{0}}}}\right)} \nonumber \\ 
 & {{s}}{{.t}}{\rm{.}}\;\;\;\text{      }{\sum\nolimits_{m=1}^{M}({{{\alpha }_{m,n}}Tr({\bm{V}_{t}}{\bm{R}_{t}})+W{{\xi }_{0}}}})\leq \frac{{{P}_{t}}|{{h}_{n}}{{|}^{2}}}{\tilde{\gamma }_{c}}, \\ 
 &\;\;\;\;\;\;\; \;\;\;{{\left[ {\bm{V}_{r}} \right]}_{l,l}}+{{\left[ {\bm{V}_{t}} \right]}_{l,l}}=1,{{\left[ {\bm{V}_{x}} \right]}_{l,l}}\ge 0,\forall l\in \{1,L\},  \\ 
 & \;\;\;\;\;\;\;\;\;\;\;\;\;\;\;\;\; {{\left[ {\bm{V}_{r}} \right]}_{L+1,L+1}}=1,{\bm{V}_{r}}\ge 0,{\bm{V}_{t}}\ge 0,\\
&\;\;\;\;\;\;\;\;\;\;\;\;\;\;\;\;\;\;\;\;\;\; rank({\bm{V}_{r}})=rank({\bm{V}_{t}})=1.
\end{align}
\end{subequations}

\begin{algorithm}[t]
\small
	\caption{The proposed AIOA framework}\label{AIOA}
	\KwData{$p_{max}$, $h_{k,n}$, $h_{n,d}$, $\textbf{h}_{k,r}$, $\textbf{h}_{r,n}$ and ${\Gamma _{{\rm{min}}}};$}
	\KwResult{${\mathbf{\bm{\rho}^{(*)}}}$, ${\mathbf{\bm{\alpha}^{(*)}}}$, ${\mathbf{\bm{\Phi}^{(*)}}}$ $;$}
	Initialize   ${\mathbf{\bm{\alpha}^{(0)}}}$, ${\mathbf{\bm{\Phi}^{(0)}}}$, $k=0$\;
	\While {$\sum\limits_{m\in \mathcal{M}}{\left( S_{m}^{(k+1)}-S_{m}^{(k)} \right)/}S_{m}^{(k+1)}\ge \delta$ }{
		Update ${\mathbf{\bm{\rho}^{(k+1)}}}$\; 
		Solve the problem (\ref{alpha}) by using CVX for given \{${\mathbf{\bm{\rho}^{(k+1)}}}$, ${\mathbf{\bm{\alpha}^{(k)}}}$\} and obtain the solution \{${\mathbf{\bm{\rho}^{(k+1)}}}$,${\mathbf{\bm{\alpha}^{(k+1)}}}$, ${\mathbf{\bm{\Phi}^{(k)}}}$\}\;
		For the given \{${\mathbf{\bm{\rho}^{(k+1)}}}$, ${\mathbf{\bm{\alpha}^{(k+1)}}}$, ${\mathbf{\bm{\Phi}^{(k)}}}$\}, solve the problem (\ref{problem36}) to obtain the \{${\mathbf{\bm{\Phi}^{(k+1)}}}$ \}\;
		{$k=k+1$;}	  
    }
\end{algorithm}
Since the constraint (\ref{problem36}c) is still not convex, we first relax the rank-first constraint and then it becomes convex. Since this is a convex semi-definite programming (SDP), here we use the standard convex optimization solver CVX, and then apply Gaussian randomization process for recover $\bm{\Theta}$ and obtain the corresponding $\bm{\Theta}^{\rm opt}_{x}$. Then, we obtain $\bm{\Phi} _{x}^{\rm opt}\!=\!diag\left( \bm{\Theta} _{x}^{\rm opt} \right)$. By alternately optimizing the above three variables, we propose an $\textbf{A}$lternative $\textbf{I}$terative $\textbf{O}$ptimization $\textbf{A}$lgorithm (\textbf{AIOA}) framework, shown in \textbf{Algorithm~\ref{AIOA}}.



\begin{figure*}[t] 
\centering
\captionsetup{font={footnotesize }}
\subfigure[]{
\includegraphics[width=0.65\columnwidth]{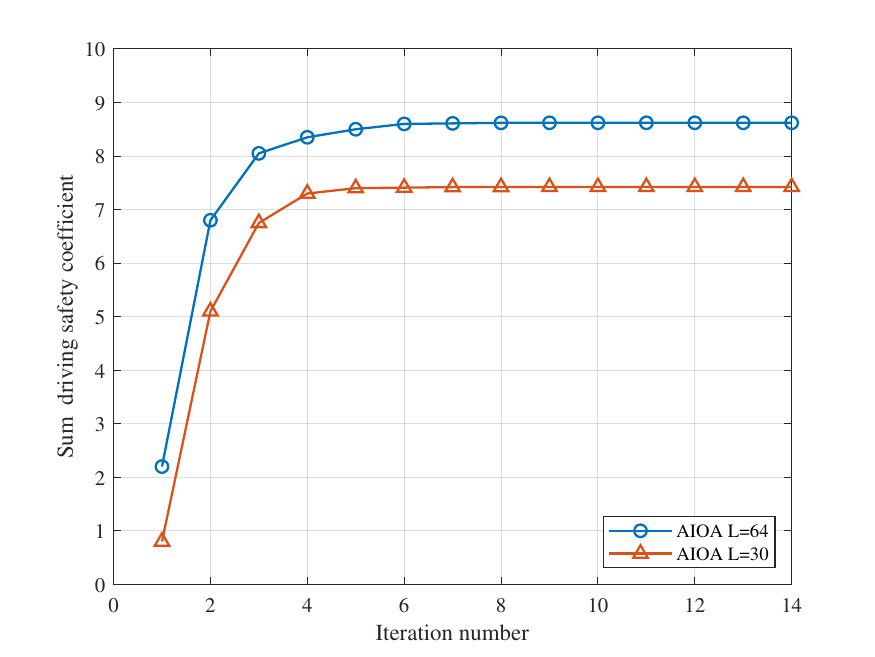}
}
\hspace{-2.5mm}
\subfigure[]{
\includegraphics[width=0.65\columnwidth]{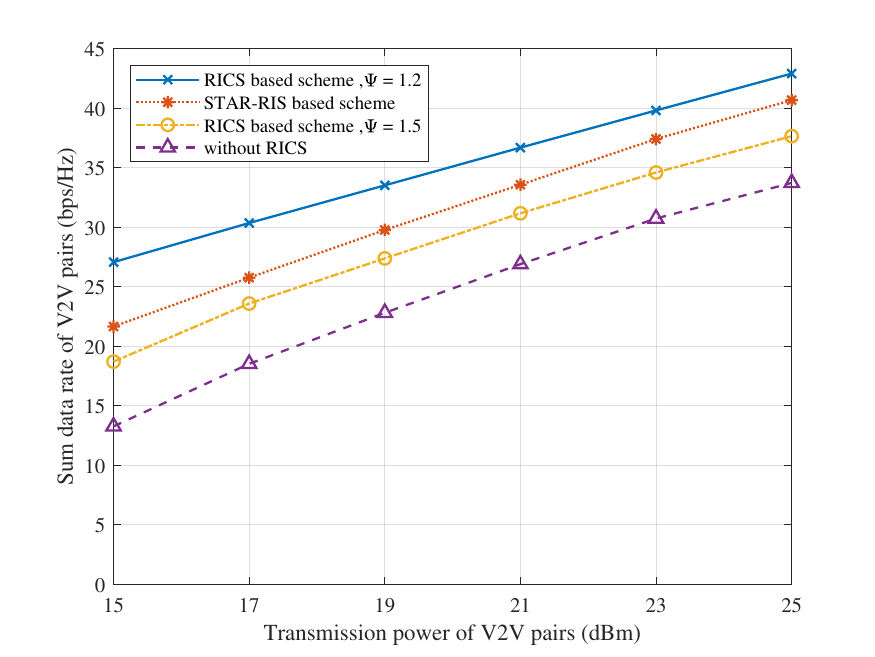}}
\hspace{-2.5mm}
\subfigure[]{
\includegraphics[width=0.65\columnwidth]{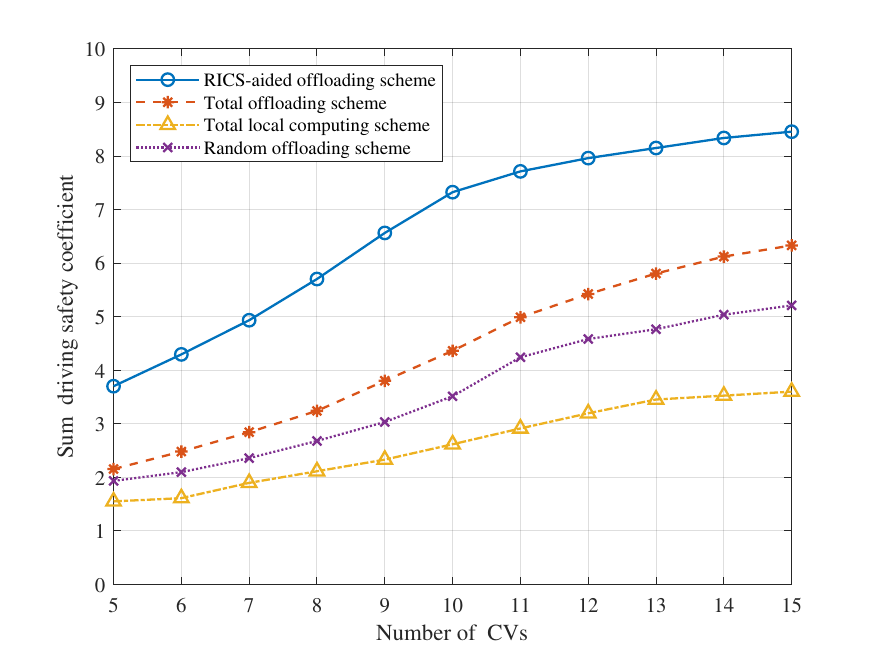}}
\caption{Convergence of the proposed AIOA algorithm is shown in (a). Available transmission power with varying value of $P_{t}$ is shown in (b), where $N=5$. Sum safety coefficient with varying number of CVs is shown in (c).}
\label{results}
\end{figure*}
\vspace{-1mm}
\section{Numerical Results}
We perform the simulations with an average of $100$ runs, in a three-dimensional spatial coordinate system, where the ${x - y}$ plane is a circular field with a radius of $400$ m and the height is $50$ m. Specifically, the CV pairs and the V2V pairs are randomly and uniformly placed. The BS is located at $\left (0, 0, 25  \right )$ and the RICS is located at $\left (100, 0, 30\right )$. Some specific simulation parameters are shown in Table~\ref{parameters}. 

\begin{table}[hbt]
	\centering
	\caption{Simulation Parameters}
	\label{parameters}
	\small
	\begin{tabular}{l | l || l | l}
		\noalign{\global\arrayrulewidth=0.3mm}
		\hline
		\textbf{Parameter}  &\textbf{Value } &\textbf{Parameter}  & \textbf{Value } \\
		\hline
		 $L$   & 30 &  $M$   &10\\
		 $N$  &10 &  $s_m$     &[10, 20] Mbits \\
		$f_m$  & [1, 5] GHz & $F$ & 10 GHz \\
		 $W{{\xi }_{0}}$    &    -110 dBm & $P_m$    &     28 dBm\\
		$P_t$ &   23 dBm & $P_{out}$        &0.01\\
		  $\delta $    &    $10^{-3}$ & $\beta $   &    4\\           
		\noalign{\global\arrayrulewidth=0.3mm}
		\hline
	\end{tabular}
\end{table}	

Fig.~\ref{results}(a) shows the convergence of the proposed AIOA algorithm with $30$ and $64$ RICS elements, respectively. It is shown that the proposed AIOA algorithm in both two cases become converged when the number of iterations is $6$. Moreover, we observe that the proposed AIOA algorithm with $L=64$ outperforms the case with $L=30$, which indicates that the average safety coefficient of each CV can reach almost $90\%$ with $64$ RICS elements. 


In Fig.~\ref{results}(b) we evaluate three additional benchmark schemes in terms of V2V pairs data rate to validate the advantages of the proposed RICS structure.  
We selected the two most representative amplification factor through extensive simulations: $\bm{\Psi}=1.2$ and $\bm{\Psi}=1.5$, respectively, to perform the effect of signal amplification factor on the data rate achieved at the V2V pairs. We observe that as the transmission power of V2V pairs increases, the achieved sum rate increases accordingly. More interestingly, it is shown that RICS-based scheme with $\bm{\Psi} = 1.2$ outperforms that with $\bm{\Psi} = 1.5$. The reason behind this is that an optimal signal amplification factor  exists, thereby helping to achieve perfect interference elimination at the $\rm Rx$ side of the V2V pair.



Fig.~\ref{results}(c) explores the influence of the number of CVs on the sum driving safety coefficient under the different offloading strategies. The three benchmark schemes are evaluated. 1) Total offloading scheme: all the $M$ CVs choose to offload their tasks to the BS.
 { 2) Total local computing scheme}: all the $M$ CVs choose to process their tasks locally.
 { 3) Random offloading scheme}: all the $M$ CVs randomly choose to offload or not.
It can be seen that our proposed RICS-aided offloading scheme outperforms the others, and after a steep rise, the sum driving safety coefficient starts to flatten out gradually when $M=10$. This is due to the fact that when $M \in [5, 10]$, the system has enough capability to bear more CVs to perform offloading while mitigating interferences suffered at the V2V pairs. Meanwhile, due to the computational limitation of the CVs, we observe that the total offloading scheme outperforms the other two benchmark schemes.

\vspace{-1mm}
\section{Conclusions}
In this paper, we presented a novel RICS-aided computation offloading framework for improving autonomous driving safety. An alternating optimization algorithm framework was illustrated to efficiently solve the formulated non-convex design problem within a few iterations. Simulation results demonstrated that the proposed RICS-aided offloading framework can not only achieve high inference accuracy of CVs' but also mitigate interference at V2V pairs.
\vspace{-1mm}


\vspace{-1mm}
\ifCLASSOPTIONcaptionsoff
  \newpage
\fi

\vspace{12pt}
\end{spacing}
\end{document}